\documentclass[12pt]{article}
\usepackage{psfrag}
\usepackage{hyperref}
\usepackage{amsmath}
\usepackage{amsthm}
\usepackage{amssymb}
\usepackage{epsfig}
\usepackage{euscript}
\usepackage{array}
\usepackage{cite}
\usepackage{cancel}
\usepackage{mathtools}
\usepackage{empheq}
\usepackage{graphics}\usepackage{subfigure}

\usepackage{hyperref}
\usepackage{verbatim}

\theoremstyle{definition}

\theoremstyle{remark}

\newcounter{multieqs}

\newcommand{\be}{\begin{equation}}
\newcommand{\ee}{\end{equation}}
\newcommand{\eq}[1]{(\ref{#1})}
\newcommand{\bit}{\begin{itemize}}  \newcommand{\eit}{\end{itemize}}

\newcommand{\bm}[1]{\mbox{\boldmath $#1$}}
\newcommand{\rf}[1]{(\ref{#1})}

\def\bd{\begin{document}}
\def\ed{\end{document}}
\def\nn{\nonumber}
\def\bea{\begin{eqnarray}}
\def\eea{\end{eqnarray}}
\let\bm=\bibitem

\def\la{\langle}
\def\ra{\rangle}

\def\npb#1#2#3{Nucl. Phys. {\bf{B#1}} #3 (#2)}
\def\plb#1#2#3{Phys. Lett. {\bf{#1B}} #3 (#2)}
\def\prl#1#2#3{Phys. Rev. Lett. {\bf{#1}} #3 (#2)}
\def\prd#1#2#3{Phys. Rev. {D \bf{#1}} #3 (#2)}
\def\cmp#1#2#3{Comm. Math. Phys. {\bf{#1}} #3 (#2)}
\def\cqg#1#2#3{Class. Quantum Grav. {\bf{#1}} #3 (#2)}
\def\nppsa#1#2#3{Nucl. Phys. B (Proc. Suppl.) {\bf{#1A}}#3 (#2)}
\def\ap#1#2#3{Ann. of Phys. {\bf{#1}} #3 (#2)}
\def\ijmp#1#2#3{Int. J. Mod. Phys. {\bf{A#1}} #3 (#2)}
\def\rmp#1#2#3{Rev. Mod. Phys. {\bf{#1}} #3 (#2)}
\def\mpla#1#2#3{Mod. Phys. Lett. {\bf A#1} #3 (#2)}
\def\jhep#1#2#3{J. High Energy Phys. {\bf #1} #3 (#2)}
\def\atmp#1#2#3{Adv. Theor. Math. Phys. {\bf #1} #3 (#2)}

\def\N{{\cal N}}
\def\sst{\scriptscriptstyle}
\def\thetabar{\bar\theta}
\def\Tr{{\rm Tr}}
\def\one{\mbox{1 \kern-.59em {\rm l}}}

\def\a{\alpha}      \def\da{{\dot\alpha}}  \def\dA{{\dot A}}
\def\b{\beta}       \def\db{{\dot\beta}}  
\def\g{\gamma}  \def\G{\Gamma}  \def\dc{{\dot\gamma}}  
\def\d{\delta}  \def\D{\Delta}  \def\ddt{\dot\delta}  
\def\e{\epsilon}        \def\ve{\varepsilon}  
\def\f{\phi}    \def\F{\Phi}    \def\vvf{\f}  
\def\h{\eta}  
\def\k{\kappa}  
\def\l{\lambda} \def\L{\Lambda}  
\def\m{\mu} \def\n{\nu}  
\def\o{\omega}  
\def\p{\pi} \def\P{\Pi}  
\def\r{\rho}  
\def\s{\sigma}  \def\S{\Sigma}  
\def\t{\tau}  
\def\th{\theta} \def\Th{\Theta} \def\vth{\vartheta}  
\def\X{\Xeta}  
\def\z{\zeta}  

\def\na{\nabla}  

\def\cA{{\cal A}} \def\cB{{\cal B}} \def\cC{{\cal C}}  
\def\cD{{\cal D}} \def\cE{{\cal E}} \def\cF{{\cal F}}  
\def\cG{{\cal G}} \def\cH{{\cal H}} \def\cI{{\cal I}}  
\def\cJ{{\cal J}} \def\cK{{\cal K}} \def\cL{{\cal L}}  
\def\cM{{\cal M}} \def\cN{{\cal N}} \def\cO{{\cal O}}  
\def\cP{{\cal P}} \def\cQ{{\cal Q}} \def\cR{{\cal R}}  
\def\cS{{\cal S}} \def\cT{{\cal T}} \def\cU{{\cal U}}  
\def\cV{{\cal V}} \def\cW{{\cal W}} \def\cX{{\cal X}}  
\def\cY{{\cal Y}} \def\cZ{{\cal Z}}

\def\ua{\underline{\alpha}}  
\def\uc{\underline{\phantom{\alpha}}\!\!\!\gamma}  
\def\um{\underline{\mu}}  
\def\ud{\underline\delta}  
\def\ue{\underline\epsilon}  
\def\una{\underline a}\def\unA{\underline A}  
\def\unb{\underline b}\def\unB{\underline B}  
\def\unc{\underline c}\def\unC{\underline C}  
\def\und{\underline d}\def\unD{\underline D}  
\def\une{\underline e}\def\unE{\underline E}  
\def\unf{\underline{\phantom{e}}\!\!\!\! f}\def\unF{\underline F}  
\def\unm{\underline m}\def\unM{\underline M}  
\def\unn{\underline n}\def\unN{\underline N}  
\def\unp{\underline{\phantom{a}}\!\!\! p}\def\unP{\underline P}  
\def\unq{\underline{\phantom{a}}\!\!\! q}  
\def\unQ{\underline{\phantom{A}}\!\!\!\! Q}  
\def\unH{\underline{H}}  
  
\def\As {{A \hspace{-6.4pt} \slash}\;}  
\def\bs {{b \hspace{-6.4pt} \slash}\;}  
\def\Ds {{D \hspace{-6.4pt} \slash}\;}
\def\Gts {{\Gt \hspace{-6.4pt} \slash}\;}
\def\ds {{\del \hspace{-6.4pt} \slash}\;}  
\def\ss {{\s \hspace{-6.4pt} \slash}\;}  
\def\ks {{ k \hspace{-6.4pt} \slash}\;}  
\def\ps {{p \hspace{-6.4pt} \slash}\;}   
\def\xs {{x \hspace{-6.4pt} \slash}\;}  
\def\pas {{{p_1} \hspace{-6.4pt} \slash}\;}  
\def\pbs {{{p_2} \hspace{-6.4pt} \slash}\;}   
\def\cFs {{{\cal F} \hspace{-6.4pt} \slash}\;}

\def\Ah{{\hat{A}}}  
\def\Dh{{\hat{D}}}
\def\Gh{{\hat{G}}}
\def\Fh{{\hat{F}}}
\def\Ih{{\hat{I}}} 
\def\Jh{{\hat{J}}} 
\def\Kh{{\hat{K}}}
\def\Lh{{\hat{L}}} 
\def\Ph{{\hat{P}}}
\def\Rh{{\hat{R}}}
\def\Vh{{\hat{V}}} 
\def\Xh{{\hat{X}}}
 
\def\ah{{\hat{\a}}}
\def\bh{{\hat{\b}}}
\def\gh{{\hat{\g}}}
\def\dh{{\hat{\d}}}
\def\hh{\hat{h}}
\def\uh{\hat{u}}  
\def\xh{\hat{x}}  
\def\yh{\hat{y}}  
\def\ph{\hat{p}}  
\def\xih{\hat{\xi}}  
\def\chih{\hat{\chi}}  
\def\Psih{\hat{\Psi}}    
\def\phih{\hat{\phi}}

\def\psit{\tilde{\psi}}  
\def\Psit{\tilde{\Psi}}   
\def\Psibt{\tilde{\bar{Psi}}}  

\def\st{\tilde{\sigma}}  

\def\delt{\tilde{\delta}}
\def\Phit{\tilde{\Phi}}   
\def\Phitb{\overline{\tilde{Phi}}}  
\def\tht{\tilde{\th}}  
\def\lt{\tilde{\l}}
\def\chit{\tilde{\chi}}   
\def\phit{\tilde{\phi}} 

\def\At{\tilde{A}}
\def\Bt{\tilde{B}}
\def\Ct{\tilde{C}}
\def\Dt{\tilde{D}}
\def\Et{\tilde{E}}
\def\Ft{\tilde{F}}
\def\Gt{\tilde{G}}
\def\Ht{\tilde{H}}
\def\It{\tilde{I}}
\def\Jt{\tilde{J}}
\def\Qt{\tilde{Q}}  
\def\Rt{\tilde{R}}  
\def\Mt{\tilde{M }}  
\def\Nt{\tilde{N}}   
\def\St{\tilde{S}}
\def\Vt{\tilde{V}}
\def\Xt{\tilde{X}} 
\def\at{\tilde{a}}
\def\ct{\tilde{c}}
\def\dt{\tilde{d}}
\def\htt{\tilde{h}} 
\def\ft{\tilde{f}}
\def\gt{\tilde{g}}
\def\pt{\tilde{p}}  
\def\qt{\tilde{q}}  
\def\vt{\tilde{v}}  
\def\nt{\tilde{n}}  
\def\ut{\tilde{u}}  
\def\wt{\tilde{w}}  
\def\zt{\tilde{z}} 
\def\xt{\tilde{x}} 
\def\yt{\tilde{y}} 
\def\Psit{\tilde{\Psi}}
\def\vphit{\tilde{\varphi}}  

\def\eb{\bar{\epsilon}} 
\def\delb{\bar{\partial}}  
\def\thb{\bar{\theta}}
\def\mub{\bar{\mu}}
\def\lamb{\bar{\l}}
\def\psib{\bar{\psi}}
\def\sb{\bar{\sigma}}
\def\xib{\bar{\xi}}
\def\chib{\bar{\chi}}

\def\Psib{\bar{\Psi}}
\def\Phib{\bar{\Phi}}
\def\Lamb{\bar{\Lambda}}
\def\Sb{{\overline \Sigma}}
\def\cb{\bar{c}}
\def\hb{\bar{h}}
\def\qb{\bar{q}}
\def\wb{\bar{w}}
\def\ub{\bar{u}}
\def\zb{{\bar{z}}}
\def\Hb{\bar{H}}
\def\Qb{{\bar Q}}
\def\Omegab{\overline{\Omega}}
\def\ob{\overline{\omega}}

\def\Ab{{\overline A}} \def\Bb{{\overline B}} \def\Cb{{\overline C}}  
\def\Db{{\overline D}} \def\Eb{{\overline E}} \def\Fb{{\overline F}}  
\def\Gb{{\overline G}} 
\def\Ib{{\overline I}}  
\def\Jb{{\overline J}} \def\Kb{{\overline K}} \def\Lb{{\overline L}}  
\def\Mb{{\overline M}} \def\Nb{{\overline N}} \def\Ob{{\overline O}}  
\def\Pb{{\overline P}}  \def\Rb{{\overline R}}  
 \def\Tb{{\overline T}} \def\Ub{{\overline U}}  
\def\Vb{{\overline V}} \def\Wb{{\overline W}} \def\Xb{{\overline X}}  
\def\Yb{{\overline Y}} \def\Zb{{\overline Z}}  

\def\fb{{\overline f}}
\def\gb{{\overline g}}
\def\mb{{\overline m}}
\def\lb{{\overline l}}
\def\yb{{\overline y}}
  
\def\ldel{{\overleftarrow{\del}}}
\def\rdel{{\overrightarrow{\del}}}
\def\ldeldel{{\overleftarrow{\del^2}}}
\def\rdeldel{{\overrightarrow{\del^2}}}
\def\ldelb{{\overleftarrow{\bar{\del}}}}
\def\rdelb{{\overrightarrow{\bar{\del}}}}

\def\ba{{\bf a}} 
\def\bk{{\bf k}}  
\def\bl{{\bf l}}  
\def\bp{{\bf p}}  
\def\bq{{\bf q}}  
\def\br{{\bf r}}
\def\bt{{\bf t}}
\def\bu{{\bf u}}
\def\bv{{\bf v}}
\def\bx{{\bf x}}  
\def\by{{\bf y}}  
\def\bR{{\bf R}}  
\def\bV{{\bf V}}

\def\bone{{\bf 1}}  

\def\va{{\vec a}}
\def\vk{{\vec k}}
\def\vp{{\vec p}}
\def\vq{{\vec q}}
\def\vx{{\vec x}}
\def\vy{{\vec y}}
\def\vu{{\vec u}}
\def\vv{{\vec v}}
\def \vH{{\vec H}}
\def \vg{{\vec g}}

\def\vs{{\vec \sigma}}
\def\vtau{{\vec \tau}}

\newcommand{\ov}[1]{\overrightarrow{#1}}

\def\frA{\mathfrak{A}}
\def\frB{\mathfrak{B}}
\def\frC{\mathfrak{C}}
\def\frD{\mathfrak{D}}
\def\frE{\mathfrak{E}}
\def\frF{\mathfrak{F}}
\def\frG{\mathfrak{G}}
\def\frH{\mathfrak{H}}
\def\frM{\mathfrak{M}}
\def\frN{\mathfrak{N}}
\def\frR{\mathfrak{R}}
\def\frW{\mathfrak{W}}

\def\fra{\mathfrak{a}}
\def\frb{\mathfrak{b}}
\def\frf{\mathfrak{f}}
\def\frg{\mathfrak{g}}
\def\frh{\mathfrak{h}}
\def\frl{\mathfrak{l}}
\def\frs{\mathfrak{s}}
\def\fri{\mathfrak{i}}
\def\frj{\mathfrak{j}}

\def\ma{\mathfrak{a}}
\def\mg{\mathfrak{g}}
\def\mh{\mathfrak{h}}
\def\mR{\mathfrak{R}}
\def\mN{\mathfrak{N}}

\def\d{\delta}\def\D{\Delta}\def\ddt{\dot\delta}  
  
\def\pa{\partial} \def\del{\partial}  
\def\xx{\times}  
\def\uno{\mbox{1 \kern-.59em {\rm l}}}    
  
\def\trp{^{\top}}  
\def\inv{^{-1}}  
\def\dag{{^{\dagger}}}  
\def\pr{^{\prime}}  
  
\def\rar{\rightarrow}  
\def\lar{\leftarrow}  
\def\lrar{\leftrightarrow}  
  
\newcommand{\0}{\,\!}      
\def\one{1\!\!1\,\,}  
\def\im{\imath}  
\def\jm{\jmath}  
  
\newcommand{\tr}{\mbox{tr}}  
\newcommand{\slsh}[1]{/ \!\!\!\! #1}  
  
\def\vac{|0\rangle}  
\def\lvac{\langle 0|}  
  
\def\hlf{\frac{1}{2}}  
\def\ove#1{\frac{1}{#1}}  

\def\Box{\square}  
\def\CC {\mathbb{C}}
\def\FF {\mathbb{F}}
\def\RR{\mathbb{R}}
\def\NN{\mathbb{N}}  
\def\ZZ{\mathbb{Z}}  
\def\bb#1{{\bf #1}}  
\def\bcomment#1{}  
\def\bfhat#1{{\bf \hat{#1}}}  
\def\VEV#1{\left\langle #1\right\rangle}  

\newcommand{\ex}[1]{{\rm e}^{#1}} \def\ii{{\rm i}}  

\newcommand{\lrbrk}[1]{\left(#1\right)}
\newcommand{\lrsbrk}[1]{\left[#1\right]}
\newcommand{\sfrac}[2]{{\textstyle\frac{#1}{#2}}}
 
\def\stw{{\sqrt{2}}}

\def\rf {{\rm f}}
\def\ri {{\rm i}}
\def\rj {{\rm j}}
\def\rn {{\rm n}}
\def\rk {{\rm k}}
\def\rl {{\rm l}}
\def\rs {{\scriptscriptstyle \rm S}}
\def\rt {{\scriptscriptstyle \rm T}}

\def\rQ {{\scriptscriptstyle \rm \cQ}}
\def\rR {{\scriptscriptstyle \rm \cR}}

\def\cQb{{\cal \Qb}}
\def\cRb{{\cal \Rb}}
\def\cWb{{\cal \Wb}}

\def\fd {{\rm N}}
\def\afd {{\overline{\rm N}}}

\def \II {I\hspace{-.1em}I\hspace{.1em}}
\def \IIA {\mbox{\II A\hspace{.2em}}}
\def \IIB {\mbox{\II B\hspace{.2em}}}
\def \gs {g^s}
\def \ls {\lambda^s}

\def \I {{\cal I}}
\def \qs {q\hspace{-.53em}/\hspace{.15em}}
\def \ks {k\hspace{-.53em}/\hspace{.15em}}
\def \YM {{\mbox{\tiny YM}}}
\def \gym {g_{\YM}}

\def \Lc {\L_c}
\def\IR{\relax{\rm I\kern-.18em R}}
\def \id {{\bf 1}}

\def\cci{\ell}
\def\ccj{\ell'}

\def \thbb{\overline{\th\th}}
\newcommand \ol{\overline}
\def \lamb{\bar{\lambda}}
\def \vphi{\varphi}
\def \lambh{\hat{\bar{\lambda}}}
\def \lh{\hat{\lambda}}
\def \dd{\ddagger}

\newcommand{\QNB}[3]{[#1,#2,#3]}
\def\hm{\tilde{\eta}} 
\def\lp{l_{+}}
\def\lm{l_{-}}
\def \PS {{(\text{PS})}}
\def \Dir {{(\text{Dirac})}}
\def \WY {{(\text{WY})}}
\def \Sin {{(\text{Sin})}}
\def \tHP{{(\text{'t-P})}}
\def \uo {{U(1)}}
\def \Lt {\tilde{L}}
\def \tn {{\tau}^{(n)}}
\def \thn {{\hat{\theta}}^{(n)}}
\def \vphin {{\hat{\vphi}}^{(n)}}

\newcommand{\Ga}{{\Gamma}}
\newcommand{\De}{{\Delta}}
\newcommand{\Lm}{{\Lambda}}
\newcommand{\Om}{{\Omega}}

\newcommand{\al}{{\alpha}}
\newcommand{\ga}{{\gamma}}
\newcommand{\de}{{\delta}}
\newcommand{\ep}{{\epsilon}}
\newcommand{\vep}{{\varepsilon}}
\newcommand{\te}{{\theta}}
\newcommand{\ka}{{\kappa}}
\newcommand{\vpi}{{\varpi}}
\newcommand{\sig}{{\sigma}}
\newcommand{\om}{{\omega}}

\newcommand{\alt}{{\rm alt}}
\newcommand{\bdy}{{\rm bdy}}
\newcommand{\bsa}{{\boldsymbol{a}}}
\newcommand{\bsb}{{\boldsymbol{b}}}
\newcommand{\bsD}{{\boldsymbol{D}}}
\newcommand{\bsk}{{\boldsymbol{k}}}
\newcommand{\bsM}{{\boldsymbol{M}}}
\newcommand{\bulk}{{\rm bulk}}
\newcommand{\cont}{{\rm cont.}}
\newcommand{\cdN}{{\mathcal{N}_d}}
\newcommand{\lan}{{\langle}}
\newcommand{\pd}{{\partial}}
\newcommand{\R}{{\rm R}}
\newcommand{\rad}{{\rm rad}}
\newcommand{\ran}{{\rangle}}
\newcommand{\Slash}[1]{{\ooalign{\hfil/\hfil\crcr$#1$}}} 
\newcommand{\srel}[2]{{\stackrel{\scriptstyle #1}{\scriptstyle #2}}}
\newcommand{\std}{{\rm std}}
\newcommand{\U}{{\rm U}} 
\newcommand{\ul}{\underline}
\newcommand{\UV}{{\rm UV}}
\newcommand{\wg}{{\wedge}}
\newcommand{\wh}{\widehat}

\def\Log{\mathop{\rm Log}}
\def\Spin{\mathop{\rm Spin}}
\def\SO{\mathop{\rm SO}}
\def\O{\mathop{\rm O}}
\def\SU{\mathop{\rm SU}}
\def\U{\mathop{\rm U}}
\def\Sp{\mathop{\rm Sp}}
\def\SL{\mathop{\rm SL}}
\def\GL{\mathop{\rm GL}}

\def\det{\mathop{\rm det}\nolimits}
\def\sign{\mathop{\rm sign}\nolimits}
\def\mod{\mathop{\rm mod}\nolimits}
\def\tr{\mathop{\rm tr}\nolimits}
\def\diag{\mathop{\rm diag}\nolimits}
\def\Re{\mathop{\rm Re}\nolimits}
\def\Im{\mathop{\rm Im}\nolimits}
\def\Tr{\mathop{\rm Tr}\nolimits}
\def\bbra{{\langle\kern-2.5pt\langle}}
\def\kket{{\rangle\kern-2.5pt\rangle}}
\def\Bbra{{\Big\langle\kern-3.5pt\Big\langle}}
\def\Kket{{\Big\rangle\kern-3.5pt\Big\rangle}}

\begin{document}
 
\hfill{DCPT-13/29}

\vspace{20pt}

\begin{center}

{\Large \bf
Non-Abelian Self-Dual Strings in Six Dimensions from Four Dimensional
$1/2$-BPS Monopoles}
\vspace{30pt}

{\bf Chong-Sun Chu}
{\em
\begin{itemize}
\item[]
Department of Physics and National Center for Theoretical Sciences, 
National Tsing Hua University, Hsinchu 30013, Taiwan. 
\item[]
Centre for Particle Theory and Department of Mathematics,
Durham University, Durham, DH1 3LE, UK.
\end{itemize}
}

\vspace{20pt}
{\bf Abstract}
\end{center} 

We explain a new 
construction of self-dual string solutions to the 
non-abelian two-form self-duality equation proposed in
\cite{CK}. 
This class of self-dual strings is 
determined by the BPS monopoles in four-dimensions and 
the self dual string charge is given by the charge of the monopole.
Our construction covers the  $SO(4)$ invariant self-dual string solutions 
found previously. 
We have also constructed, based on the
't  Hooft-Polyakov monopole, a singular 
solution that describes two finitely separated M5-branes meeting midway 
in between. 
We comment that 
as BPS monopoles are generally given by the  Nahm construction,
our construction suggests that a generalized Nahm
transform may exist for the
non-abelian self-dual strings.


\setcounter{page}0
\newpage

\section{Introduction}

The theory of $N$ coincident M5-branes in a flat spacetime is given by an 
interacting
(2,0) superconformal theory in six dimensions \cite{witten0}. 
The understanding of the dynamics
of this system is of utmost importance. It will not only improve our understanding
of the AdS/CFT correspondence for the $AdS_7\times S^4$ background \cite{malda}; 
in addition, as the problem involves a mathematical 
formulation of a self-duality equation
for a non-abelian 3-form gauge field strength, one may suspect that it may have
an impact on mathematical physics in  a way similar to it's 
lower dimensional cousin, the self-dual Yang-Mills equation
\cite{ym}.

On general grounds, the theory of multiple M5-branes does not have
a free dimensionless parameter and is inherently non-perturbative. 
It does not mean
that an action does not exist, though it does mean that the action will
be of limited use, probably no more than giving the 
corresponding equation of motion.
This is still very interesting since one can expect that  
non-trivial spacetime physics of M-theory could be learned from the physics of the 
solitonic objects of the worldvolume theory of M5-branes, much like 
the cases of M2-branes and D-branes. See for example, \cite{tong}.

In a recent paper \cite{CK}, a consistent self-duality equation of 
motion for a non-abelian tensor gauge field in six dimensions has been 
constructed and proposed to be the low energy equation of motion of the
self-dual tensor field living on the worldvolume of a system of 
multiple M5-branes. 
The self-dual equation of motion proposed in \cite{CK} is meant to be an 
effective description for the M5-branes, 
just like the supergravity equation of motion provides
an effective description for the M-theory. Recently it has been conjectured 
\cite{dou,lam2} that
the 5d supersymmetric Yang-Mills theory can be used to provide a fundamental
definition of the (2,0) theory. This conjecture has been checked quite recently
in \cite{dixon} and it was shown that the  5d supersymmetric Yang-Mills theory
is  divergent at six loops and hence extra degrees of freedom is needed to 
provide a UV completion. Having a workable fundamental
definition for the M5-branes system is highly desirable but very difficult, 
see  also \cite{qm1,qm2,dec} for other proposals. 
We will be restricting ourselves with
the effective description and hope  something useful can be learned.

The non-abelian self-duality equation constructed in \cite{CK} 
generalizes the equation of motion
for a single M5-brane of \cite{hs,PS,schw1,pst}. 
It  was constructed 
in the gauge $B_{5\m} =0$ ($\m =0,\cdots, 4$) and is a non-abelian 
generalization of the 
Henneaux-Teitelboim-Perry-Schwarz construction for the $U(1)$ case \cite{PS,HT}.
The construction of \cite{CK} involves the introduction of a set of 
non-propagating non-abelian 1-form gauge fields which was motivated originally 
by the boundary analysis  in \cite{CS} and  further analyzed
 \cite{chu}. This aspect is very similar to the BLG \cite{BLG}
and ABJM model \cite{ABJM} of multiple M2-branes where a set of
non-propagating Chern-Simons gauge fields was introduced in order to
allow for a simple representation of
the highly non-linear and non-local self interactions of the matter fields
of the theory.

The proposed self-duality equation  reads
\be\label{sd-na}
\Ht_{\m\n} = \pa_5 B_{\m\n},
\ee
where the gauge field $A_\m$ is constrained to be  given by
\be \label{FH}
F_{\m\n} = c\int dx_5 \Ht_{\m\n}.
\ee
Here
\be
H_{\m\n\r}= D_{[\m}B_{\n\r]} = \pa_{[\m}B_{\n\r]}+[A_{[\m},B_{\n\r]}],
\ee
\be
\tilde{H}_{\m\n} = \ove{6}\e_{\m\n\r\s\t}H^{\r\s\t},\qquad \e_{01234}=-1,
\ee
\be
F_{\m\n} = \pa_{\m} A_{\n} - \pa_{\n} A_{\m} + [A_\m,A_\n].
\ee
All fields are in the adjoint representation of the Lie algebra of 
the gauge group $G$, and $c$ is a free parameter. 
Our convention for the Lie algebra are:
$[T^a, T^b] = i f^{abc} T^c$, 
$F_{\m\n}= i F_{\m\n}^a T^a$, $A_\m = i A_\m^a T^a$ and
$F_{\m\n}^a = \del_\m A_\n^a - \del_\n A_\m^a  - f^{abc} A^b_\m A^c_\n$.

Evidence that this self-duality equation describes
the physics of multiple M5-branes was provided in \cite{CK}, and further
in \cite{CKV,CV,CI}.
In
\cite{CKV,CV}, non-abelian self-dual string solutions were constructed
and a precise agreement \cite{CV} of the 
field theory results and the supergravity descriptions \cite{siampos} 
was found. Moreover it was found that the constant $c$ is 
fixed by quantization condition of the self-dual strings solution 
of the theory. 
This is satisfying  as otherwise $c$ would be 
a free dimensionless constant in the theory and hence contradicts with
what we know about M5-branes in flat space.
In \cite{CI}, non-abelian wave configurations 
which are supported by Yang-Mills instanton were constructed
and they were found to match up nicely with the description of 
M-wave on the worldvolume of M5-branes system.

One thing interesting about the self-dual string 
solutions constructed in \cite{CKV,CV} is that the auxiliary gauge field
is always given by a magnetic monopole  which gives rise
to the charge of the self-dual string.
This was shown to be case for the original 
Perry-Schwarz self-dual string and the Wu-Yang self-dual string \cite{CKV},
as well as for the generalized Wu-Yang self-dual string \cite{CV}, with the
corresponding monopole configurations given by the Dirac monopole, 
the Wu-Yang monopole and the generalized Wu-Yang monopole. It is natural
to ask if this connection with monopole is a general feature of 
the non-abelian self-dual string. In this paper, we show that this 
is indeed the case. This result is potentially interesting as, 
given this rather explicit connection between BPS monopole 
and self-dual string, 
one may be able to provide a Nahm like
construction for non-abelian self-dual string, which has been 
speculated and analyzed by other authors \cite{saemann}. 

In the next section, we provide a general 
formalism for the construction of non-abelian
self-dual string starting from an $1/2$-BPS monopole solution in  four dimensions. 
In section 3, we show that one can recover the previously constructed
self-dual string solutions with this new formalism. 
We also construct, for the  't Hooft-Polyakov 
monopole \cite{hooft,polyakov},  a singular solution that describes two finitely
separated M5-branes meeting midway in between.
The paper is concluded with some 
further discussions in section 4.

\section{A General Construction of Self-Dual Strings in terms of BPS Monopoles}
 
In this section, we give a general construction for 
self-dual strings solutions to the non-abelian self-duality equations 
\eq{sd-na}, \eq{FH}. We will be interested in 
static configurations with  the self-dual string being infinite 
long straight line, say, in the $x^4$-direction. 
As a result, 
physical properties of the system 
are independent of $x^0$ and $x^4$.

Let us consider an ansatz 
with the following non-vanishing components of the
$B$-field:
\be 
B_{ij}, \qquad B_{04} := -   \phi.
\ee
The non-vanishing components of $H$ are
\bea
H_{ijk} = D_{[i} B_{jk]},\qquad && H_{5ij} = \del_5 B_{ij}, \label{sdh1} \\
H_{04i} = -D_i \phi, \qquad && H_{045} = -\del_5 \phi, \label{sdh2}
\eea
and the self-duality equation \eq{sd-na} reads
\bea
 \del_5 \phi &=& - \frac{1}{2} \e_{ijk} D_i B_{jk},\label{sd1} \\
D_k \phi &=& \frac{1}{2} \e_{kij} \del_5 B_{ij}.  \label{sd2}
\eea

We remark that for $U(1)$ gauge group, the self-duality 
equations \eq{sd1}, \eq{sd2} are precisely the
same as the BPS equations of Howe-Lambert-West \cite{HLW} for the (2,0)
M5-branes theory in the case when 
only a single worldvolume scalar field  $\phi:=\phi^6$
is turned on.
In \cite{CKV,CV}, it was postulated that \eq{sdh1}, \eq{sdh2} 
are the BPS equation for the  non-abelian (2,0) theory.

\subsection{General construction}

The equation \eq{sd2} 
can be integrated and solved by 
\be \label{BDphi}
B_{ij} = \e_{ijk} D_k \Phi, 
\ee
where $\Phi$ is related to $\phi$ by
\be \label{pP}
\phi =\del_5 \Phi.
\ee
Substitute \eq{BDphi} into \eq{sd1}, we obtain immediately
the 4-dimensional covariant Laplace equation
\be \label{cov-lap}
(\del_5^2 + D_i^2 ) \Phi =0.
\ee

The constraint \eq{FH} now reads
\bea 
&& F_{ij} =  c \, \e_{ijk} D_k \Phit, \label{FH1} \\
&& F_{04} =  -c ( \phi(x^5=\infty) - \phi(x^5=-\infty)), \label{FH2}
\eea
where
\be \label{bc0}
\Phit (x^i) :=  \Phi(x^i, \infty) - \Phi(x^i,-\infty).
\ee
The equation \eq{FH1} can be solved immediately by noticing 
that is takes precisely
the form of the BPS equation of a magnetic monopole, with $c \Phit$ being the 
adjoint Higgs
scalar field of the Yang-Mills theory. We note in passing that
$\Phit$  satisfies the 3-dimensional 
covariant Laplace equation
\be \label{3d-laplace}
D_i^2 \Phit =0. 
\ee
For excellent reviews of monopole, see for example, \cite{shnir,mono1,mono2}.
As for solving \eq{FH1}, we can distinguish two cases. 
For a regular solution $\Phi(x^i, x^5)$ with well defined limits 
$\Phi(x^i, \pm \infty)$, we have 
\be \label{p5p5}
\phi (x^5=\infty) = \phi (x^5=-\infty) =0.
\ee 
In this case the constraint
\eq{FH2} can be solved conveniently by having $A_0=A_4=0$. 
We will also be interested in solutions with singularities such that
non-vanishing values of $\phi (x^5=\pm\infty)$ are allowed. In this case, the
constraint \eq{FH2} is solved by
$ A_4=0, A_0 =  c ( \phi(x^5=\infty) - \phi(x^5=-\infty)) x^4$.

Base on the above observation, 
a general self-dual string
solution of the self-duality equations \eq{sd1} and \eq{sd2} can
be constructed 
entirely in terms of the solution of the   
covariant Laplace equation \eq{cov-lap} for an adjoint scalar field.
Our algorithm is to start with  
{\it any} BPS monopole configuration  $(A_i, \Phi^{(0)})$ of the 
Yang-Mills Higgs system as a seed, and looks for a solution of 
the covariant Laplace equation
\eq{cov-lap} with the boundary condition 
\be \label{bc} 
c \Phit (x^i) = \Phi^{(0)}(x^i),
\ee
where $\Phit(x^i)$ is given by \eq{bc0}. 
Then a self-dual string 
solution solving the self-duality equations \eq{sd1}, \eq{sd2} and 
the constraints \eq{FH1}, \eq{FH2}  is 
given by
\bea
B_{ij} &=& \e_{ijk} D_k \Phi, \label{ff1} \\
\phi &=& \del_5 \Phi. \label{ff2}
\eea
In addition,  we also need to specify certain boundary conditions on $\phi$ (see
\eq{bct} below)
so that the  field strengths decrease 
fast enough at infinity in order for the charge to be well defined.
This will be examined next.



\subsection{Charges}

A general feature of the non-abelian monopole is that
the gauge symmetry $G$ is broken down 
asymptotically to a little group $H$ 
by the large $r$ values of the scalar field $\Phit$.
$H$ generally contains a $U(1)$ factor generated by 
\be
T:=   T^a \hat{n}^a, 
\ee
where the unit vector $\hat{n}^a$ is determined by the asymptotic configuration 
of $\Phit$. 
As a result, our non-abelian self-dual string solution inherits  
at large $r$ (and for any $x^5$) the same unbroken gauge symmetry.
This allows us to define an asymptotic $U(1)$ field  
by a projection
\be
B_{\m\n} := B^a_{\m\n} \hat{n}^a.
\ee
It has the asymptotic $U(1)$ field strength 
\be
H_{\m\n\l} := \del_{[\m} B_{\n\l]} = D_{[\m} B_{\n\l]} {}^a \hat{n}^a.
\ee
The magnetic and electric charges (per unit length) of our self-dual string
solution is then given by
\be
P = Q = \frac{1}{2\pi^2} \int_S  H,
\ee
where $S$ is the boundary surface of a large volume $V$ in  
$\bR^4$ (of $x^1, x^2, x^3, x^5$)  
containing the self-dual string configuration.
We claim that the self-dual string charge is indeed given by 
charge of the seed BPS monopole
\be \label{PQ-claim}
P =Q = \frac{c}{2\pi^2}\int_{S^2} F,
\ee
where $S^2$ is a 2-sphere in the space of $x^1, x^2, x^3$ containing 
the BPS monopole and  
$F := F^a \hat{n}^a$ is the asymptotic $U(1)$ field strength for the 
non-abelian monopole.

To see this, let us take $V$ to be the hypercube with
boundary defined by the eight $\bR^3$-hyperplanes:
\be
S: \quad x^5 = \pm \infty, \quad \mbox{or} \quad x^i = \pm \infty.
\ee
This gives
\be
2 \pi^2 P = 2\pi^2 Q = I_5 +I_1 +I_2 +I_3,
\ee
where
\be
I_5 := \int_{\bR^3}  H_{123} \Big|_{x^5 =-\infty}^{x^5 =\infty}  dx^1 dx^2 dx^3, 
\ee
and
\be \label{I1} 
I_1 := \int_{\bR^3} H_{5 23} \Big|_{x^1 =-\infty}^{x^1 =\infty}  dx^5
dx^2 dx^3,
\ee
with $I_2$ and $I_3$ defined similarly.
To evaluate the integrals,
we compute the field strengths.
We have
\bea
H_{123} \Big|_{x^5 =-\infty}^{x^5 =\infty} &=& 
\Big[\del_1 (B^a_{23}(x^5=\infty)-B^a_{23}(x^5=-\infty)) 
+ \mbox{(123 cyclic)} \Big]
 \hat{n}^a  \nn\\
&=& c \Big[ \del_1 F^a_{23} + \mbox{(123 cyclic)} \Big] \hat{n}^a \nn\\
&=& c ( \del_1 F_{23} + \mbox{(123 cyclic)} ), \label{Hh}
\eea
where we have used \eq{FH} and
$F := F^a \hat{n}^a$ is the asymptotic $U(1)$ field strength for the 
non-abelian monopole. Therefore \eq{Hh}
is non-vanishing at the position of the monopoles and we have
\be
I_5 = \frac{c}{2\pi^2}\int_{S^2} F.
\ee
As for the other integrals $I_1, I_2, I_3$, 
we require that 
\be
H_{523} = D_1 \phi^a \; \hat{n}^a \to 0 \quad \mbox{as $x^1 \to \pm \infty$}
\ee
tends to zero fast
enough so that the integral $I_1$ vanishes.  Similar considerations apply
for $I_2$ and $I_3$. In total, we require that
\be \label{bct}
D_i \phi^a \;  \to 0 \quad \mbox{as $x^i \to \pm \infty$}.
\ee
As a result, we obtain \eq{PQ-claim} as claimed.

\section{Examples}

In this section, we give explicit 
examples to illustrate our general construction. We first show that  the
known solutions such as the Perry-Schwarz self-dual string \cite{PS} for the
$U(1)$ theory
and the Wu-Yang monopole string \cite{CKV,CV} for the non-abelian theory 
can both be
considered as examples of our general construction. 
We  then consider the 't  Hooft-Polyakov BPS monopole and show that
no acceptable self-dual string solution can be constructed. And we discuss the
possible physical reason behind this negative result. 

\subsection{Perry-Schwarz self-dual string 
and Wu-Yang monopole self-dual string}

The Perry-Schwarz self-dual string solution is given by the following configuration
of $B$-field:
\bea
B^{\rm (PS)}_{ij} &=& - \frac{\beta}{2} \frac{\e_{ijk}x^k}{r^3}
\Big[ \frac{x^5 r}{\rho^2} + \tan^{-1} (\frac{x^5}{r})\Big], \\
B^{\rm (PS)}_{04} &=& -\frac{\b}{2 \rho^2},
\eea
where $r^2 = (x^i)^2$ and $\rho^2 =(x^i)^2  +(x^5)^2$.
This is not the original form Perry-Schwarz obtained, but, as was shown
in \cite{CKV}, is gauge equivalent to it. In this Abelian case, $A_i$ decoupled
from the Laplace equation \eq{cov-lap} and so we have to solve
\be
(\del_5^2 + \del_i^2) \Phi =0.
\ee
The 4-dimensional Laplace equation has the solution $1/\rho^2$. But this has 
trivial boundary condition. It is easy to see that it's integral
\be \label{Phi-PS}
\Phi := \int^{x^5}_{-\infty} \frac{\b}{2 \rho^2}  dy^5 = 
\frac{\b}{2 r} \Big[\tan^{-1} (\frac{x^5}{r})+ \frac{\pi}{2}\Big]
\ee
also satisfies the Laplace equation \eq{cov-lap}. Moreover, it satisfies
the boundary condition \eq{bc} with
\be
\Phit = \frac{\b \pi}{2 r} .
\ee
This gives 
\be
b_i := \frac{1}{2}\e_{ijk}F_{jk} =\del_i (c \Phit) = 
-\frac{c\b\pi}{2} \frac{x^i}{r^3}.  
\ee
This is precisely the magnetic field strength for a Dirac monopole of charge $n$
if
\be
\b = -\frac{2n}{c\pi}.
\ee
Our \eq{Phi-PS} reproduces precisely the Perry-Schwarz $B_{ij}$. By construction
\be
\phi = \frac{\b}{2 \rho^2} +v.
\ee

In the paper \cite{CKV}, a self-dual string solution for the $SU(2)$ theory 
is  obtained with its auxiliary Yang-Mills field given by the generalized Wu-Yang
monopole configuration:
\be
A_i^a = -\e_{aik} \frac{x^k}{r^2}.
\ee
It is easy to see that for the ansatz, 
\be
\Phi = \frac{i x^a T^a}{r}  \vphi,
\ee 
where $T^a$, $a=1,2,3$, are the $SU(2)$ generators,
\be
[T^a,T^b] = i \e^{abc} T^c.
\ee
The covariant Laplace equation simplifies to
\be
(\del_5^2+ \del_i^2) \vphi =0.
\ee  
We can take $\vphi$ to be given by \eq{Phi-PS}. This gives
\be
\Phit =  \frac{i x^a T^a}{r} \vphit, \quad \mbox{where}\quad  
\vphit:= \frac{\b \pi}{2r},
\ee
\be
F_{ij} = \e_{ijk} D_k (c\Phit) = -\frac{c\b\pi}{2} \frac{\e_{ijk} x^k}{r^3}
\, \frac{i x^a T^a}{r}
\ee
and
\bea
B_{ij} = \e_{ijk} D_k \Phi = B_{ij}^{\rm (PS)} \frac{i x^a T^a}{r}, \nn\\
B_{04} = -\del_5 \Phi = B_{04}^{\rm (PS)} \frac{i x^a T^a}{r} .
\eea
By construction
\be
\phi = \frac{i x^a T^a}{r} (\frac{\b}{2 \rho^2} +v). 
\ee
Similarly, one can reproduce the generalized Wu-Yang monopole string obtained
in \cite{CV}. 
Note that, up to a gauge transformation, the
scalar $\phi$ as well as the field strength  are $SO(4)$ rotational 
invariant.

\subsection{'t Hooft-Polyakov monopole self-dual strings}

Next we consider the 't Hooft-Polyakov monopole. 
The 't Hooft-Polyakov monopole solution is a topological soliton of the
non-abelian $SU(2)$ gauge  
Yang-Mills gauge theory which is non-singular and carry a magnetic 
charge
since it has an asymptotic behavior similar to that of a Dirac monopole. 
In the BPS limit, the 't Hooft-Polyakov monopole satisfies the first-order BPS 
equation
\be
F_{ij} = \e_{ijk} D_k \Phi^{\rm (HP)}
\ee
and has an analytic solution  given in terms of elementary 
functions $k(r)$, $h(r)$ as
\be \label{bps-mono}
A_0^a=0, \qquad 
A_i^a = -\e_{aik} \frac{x^k}{r^2}(1-k(r)), 
\ee
\be
\Phi^{\rm (HP)}{}^a =\frac{x^a}{r^2} h(r),
\ee
where 
\be
k(r) = \frac{r}{\sinh r}, \qquad  h(r) = r \coth r -1.
\ee
At large $r$, it is
\be
k \to 0, \quad h \to r, \qquad \mbox{for $r \to \infty$}.
\ee
This means the scalar field approaches asymptotically 
a constant vacuum expectation value with the magnitude ($|B|:=\sqrt{B^a B^a}$)
\be
|\Phi^{\rm (HP)}| \to 1
\ee
and an unbroken $U(1)$ gauge symmetry can be identified there by 
projection. We have taken, for simplicity, the vacuum expectation value 
of the scalar field to be equal to 1.
The unbroken $U(1)$ is generated by the generator
\be
T = \frac{i x^a T^a}{r}
\ee
and the asymptotic $U(1)$ field strength is precisely equal to that of the 
Dirac monopole of unit charge
\be
F := F^a \frac{i x^a}{r} = F^{(\rm Dirac)}
\ee
and so the 't  Hooft-Polyakov monopole carries a unit monopole charge
\be
P = \frac{1}{2\pi} \int_{S^2} F = 1.
\ee
Despite the existence of a non-zero charge, the solution is regular everywhere.
In fact, for small $r$,
\be
k \to 1, \quad h \to -1
\ee
and so 
\be
A_\mu^a \to 0, \quad  |\Phi^{\rm (HP)}| \to 0.
\ee

Now, let us start with the BPS monopole configuration \eq{bps-mono}
and apply the general construction in the last section. As the 
't  Hooft-Polyakov monopole is  radial symmetric, the resulting self-dual string 
should be at least $SO(3)$ invariant. This implies that $\Phi$ should be of the 
form
\be \label{an1}
\Phi  = \frac{i x^a T^a}{r} \vphi(x^5, r).
\ee
The covariant Laplace equation \eq{cov-lap} then reduce 
to the single partial differential equation on $\vphi$
\be \label{de-vphi}
\Big(\del_i^2 + \del_5^2 - \frac{2 k^2(r)}{r^2} \Big) \vphi =0.
\ee
Substituting
\be \label{an2}
\vphi = \frac{H(r,x^5)}{r}
\ee
and the equation \eq{de-vphi} is further reduced to
\be \label{f-eqn}
H'' + \ddot{H} - \frac{2}{\sinh^2 r} H =0,
\ee
where $H' = \del_r H$ and $\dot{H} = \del_5 H$.
The differential equation is solved generally by
\be \label{an3}
H = \coth r \times f - \frac{\del f}{\del r},
\ee
provided $f = f(x^5,r)$ satisfies the 2-dimensional Laplace equation
\be \label{de-f}
(\del_r^2 + \del_5^2) f =0.
\ee

General solution to the Laplace equation \eq{de-f} can be readily 
written down using the theory of complex variables. 
The less trivial part is to find solution so that the boundary 
conditions \eq{bc} and \eq{bct} are satisfied.
To satisfy \eq{bc}, it is 
needed that
\be \label{bc-hp}
c(f(x^5 =\infty) - f(x^5 =-\infty))  = r,
\ee 
Let us first consider solution that is finite at infinite $x^5$, corresponds
to the situation of \eq{p5p5}. Such 
regular solution can be represented by the Fourier series 
\be
cf= \int_0^1 dp\; \a(p) \frac{\sin (p x^5)}{p} \frac{\sinh(p r)}{p}, 
\ee
without including the linear term in $x^5$. 
It is easy to see that the condition \eq{bc-hp} is satisfied with $\a(0) = c/\pi$ 
since $\frac{1}{p} \sin(p x^5) \to \pm \frac{\pi}{2} \d(p)$ 
as $x^5 \to \pm \infty$. 
This solution however blows up ($\sim e^r$) at large $r$ and is nonphysical
for a self-dual string. 
For our problem at hand, 
the most general solution to \eq{de-f} is of the form
\be \label{fs}
cf= \int_0^1 dp\; \big(
\a(p) \frac{\sin (p x^5)}{p} e^{-p r} + \b(p) \cos (p x^5) e^{-p r}
\big),
\ee
where we have not included terms $e^{p r}$ for $p>0$, and without loss
of generality, we have taken the upper limit of integration to be 1. Noticing that
$\cos(p x^5) \to 0$ as $x^5 \to \pm \infty$, we see that 
it is impossible to satisfy \eq{bc} for any choice of the 
coefficients $\a(p)$ and $\b(p)$ . 

Let us try to understand the physical reasons for 
the absence of self-dual string solution in this case.
According to our general analysis in section 2.2, 
the charge of the self-dual string is given by the charge of the BPS monopole, 
which for the case of  't Hooft-Polyakov monopole is one. As we mentioned above,
it is quite suggestive that 
the self-duality equation \eq{sd1}, \eq{sd2} coincides with the BPS equation 
of the (2,0) theory when only  a single scalar field $\phi$ is turned on. If this 
conjecture is indeed correct, then the self-dual string solution we were seeking for 
should be $SO(4)$ rotational invariant as
it should be a minimal energy configuration (being supersymmetric) 
and has a single center (having unit charge). 
However 
an 
SO(4) invariant solution 
corresponding to a unit charge self-dual string has already been constructed 
previously for general $SU(N_5$) gauge group \cite{CV}. 
This solution was based on the Wu-Yang monopole and the scalar profile 
of this solution has been found to match precisely with the 
radius-transverse distance relation obtained from a
supergravity analysis of the M2-brane spike intersecting  a system of 
M5-branes \cite{siampos}. 
It would be puzzling if one is able to find a different field theory
solution that describes an M2-brane ending on a system of M5-brane.

Physically, there are other interesting configurations one may have for a system
of  M2-branes and M5-branes. For the analogous system of D-branes,
one interesting configuration is to have a fundamental string or D-string 
stretching between two D3-branes as described by the 
wormhole like solution constructed by Callan-Maldacena \cite{CM}
and Gibbons \cite{gibbons} (see figure 1a). 
This configuration is, however,  
non-BPS as the two D3-branes are smoothly connected with
each other through the throat and hence are oppositely orientated. 
Another interesting configuration considered by Hashimoto \cite{hash} is for the two 
D3-branes to meet at precisely the point where the throat size $r$ shrinks to zero
(see figure 1b). 
The configuration is BPS as the two D3-branes are connected through a singularity at
$r=0$ and is described by the 't Hooft-Polyakov BPS monopole 
in the $SU(2)$ SYM theory. 

\begin{figure}[htb]
\begin{center}
\includegraphics[height=2in]{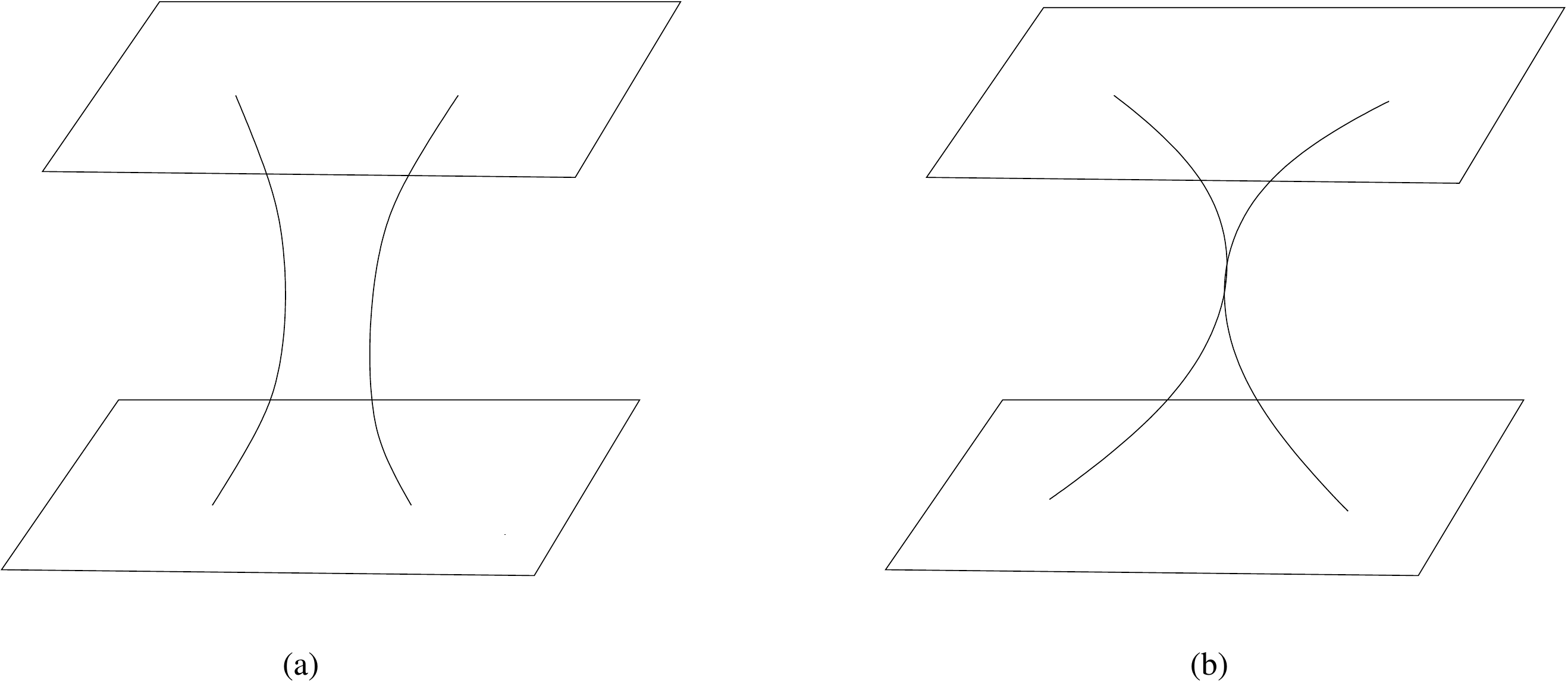}
\caption{A string stretches between a pair of parallel D3-branes.
(a) is described by a wormhole like solution. (b) is singular at $r=0$ 
and is described by a BPS monopole. }
\end{center}
\end{figure}

Back to our problem of finding solution of the equations \eq{sd1}, \eq{sd2}.  
In addition to the spike solution \cite{CKV,CV} which describes 
 a bunch of M2-branes ending on a system of
M5-branes,
we wish to find a configuration similar to the one of Hashimoto for D-branes
\footnote{
Obviously we do not expect to find a solution to 
our equations \eq{sd1}, \eq{sd2} to describe an M2-brane suspended 
between two M5-branes as this configuration is non-BPS.}. 
To get  this solution, 
we will have to relax ourself to consider solution with singularity.
So far we have considered solution that is finite at infinite $x^5$. 
In order to describe the presence of two M5-branes, in view of \eq{p5p5},
one has to give up the requirement of $\Phi$ being finite at infinite $x^5$. 
Therefore 
let us relax the regularity condition of $\Phi$ 
at  infinite $x^5$ and consider a solution of the form
\be
f := \begin{cases}
x^5 r +r/c, \quad x^5 \geq 0,\\
- x^5 r , \qquad x^5 <0. 
\end{cases}
\ee
We could also add to it a Fourier series part
but this would not affect the discussion below.
It is clear that the corresponding $\Phi$ is  discontinuous at $x^5=0$. 
Also it is
singular at infinite $x^5$. To regular the singularity, we can put the system in 
a box of size $L$: $-L \leq x^5 \leq L$. Then the boundary condition \eq{bc-hp}
is satisfied. One can check the boundary condition \eq{bct} is also satisfied. 
This solution looks rather trivial. What does it correspond to physically?
To see this, note that
\be
\phi =  \Phi^{\rm (HP)}(r) \times
\begin{cases} 1, \qquad x^5 \geq 0,\\
-1, \qquad x^5 <0. 
\end{cases}
\ee
We could diagonalize this field so that its eigenvalues can be interpreted 
as the positions of the M5-branes. As is well known from the analysis 
of the 't Hooft-Polyakov monopole, this 
can be achieved locally with an appropriate gauge transformation and we find
\be
\phi =  \pm (\coth r -\frac{1}{r})  \times
\begin{cases} 1, \qquad x^5 \geq 0,\\
-1, \qquad x^5 <0. 
\end{cases}
\ee
This solution is illustrated in figure 2.
Asymptotically as $r \to \infty$, $\phi$ approaches the constant values $\pm 1$,
corresponds to having two M5-branes placing at the constant positions 
$\phi =\pm 1$.
Near $r=0$,   $|\phi| \sim r/3$, meaning the cross-section of the 
M5-brane shrink linearly to zero as one approaches the point
halfway between the two M5-branes. This solution describes a pair of parallel
M5-branes with their respective M2-brane spikes meeting at halfway between the 
two M5-branes.

\begin{figure}[htb]
\begin{center}
\includegraphics[height=2in]{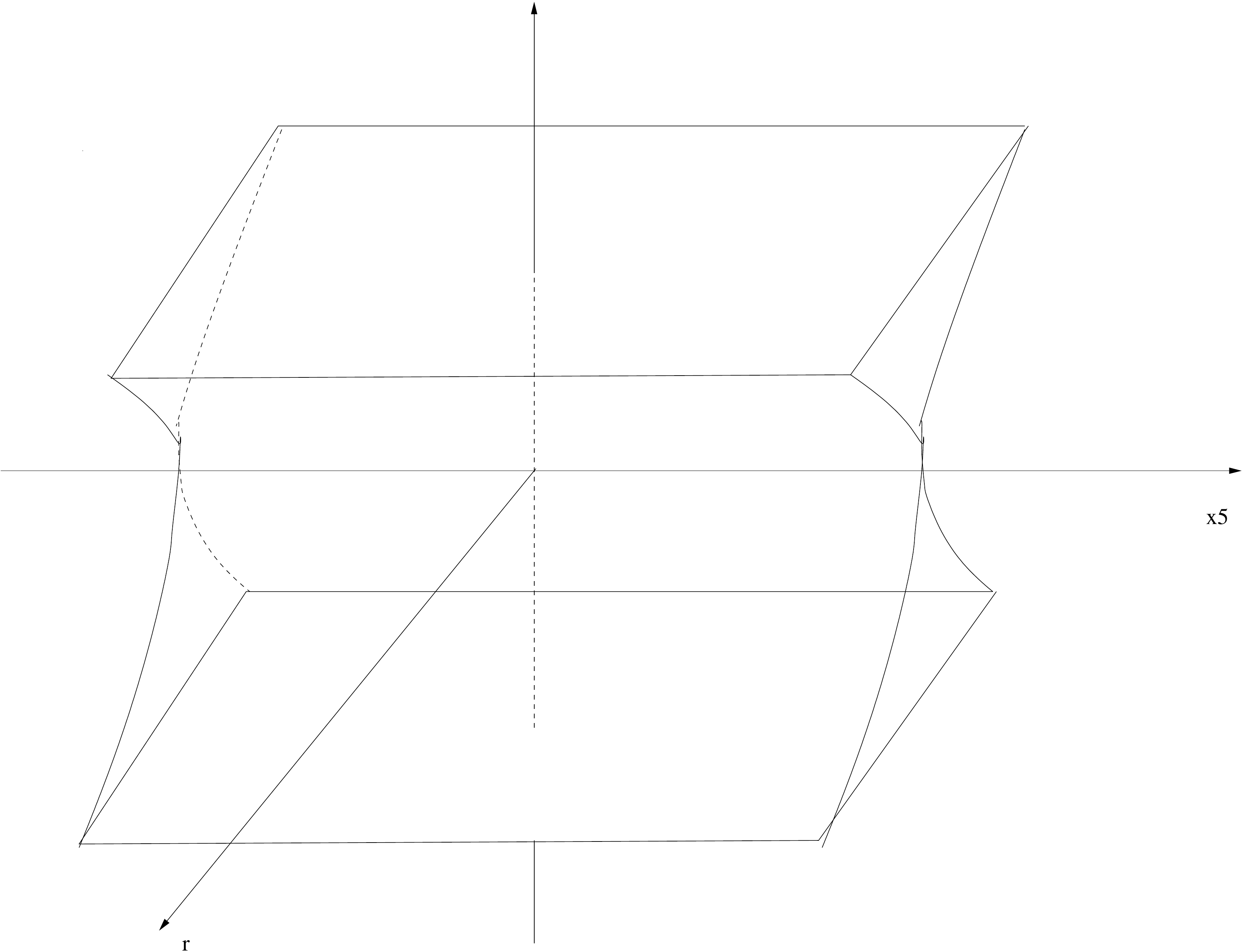}
\caption{Two parallel M5-branes meeting midway in between. }
\end{center}
\end{figure}

\section{Discussions}

In this paper, we have given a general formalism for 
the construction of  a class of non-abelian 
self-dual string solution whose auxiliary gauge field is given by a  
BPS magnetic monopole. The charge of the
self-dual string is given, up to a proportional constant, by the charge 
of the monopole. We have shown that previously found $SO(4)$ invariant solution
can be recovered by this formalism. We have also constructed, based on the
't  Hooft-Polyakov monopole, a singular 
solution that describes two finitely separated M5-branes meeting midway 
in between.


We remark that a class of string solitons of the    
models of \cite{SSW} has been found in the papers \cite{g1,g2}. 
These solutions are smooth and have charges 
supported by instanton configurations. This is to be compared with the
behavior of the string solution obtained in \cite{PS,HLW}, and those
obtained in \cite{CKV,CV}
and here, which are singular at the position of the string.  
A better understanding of these solutions and their differences 
will help us to understand better the
differences of these models.

Having analyzed the charge one sector, it 
is natural to consider the higher charges case. 
Rather explicit expressions exist for the charge two monopoles 
\cite{ward} and the corresponding self-dual string solution is expected to 
have two centers and hence $SO(3)$ invariant. 
It would be interesting to employ the present formalism to construct 
this self-dual string solution.

For higher charges, the BPS monopole configurations are given systematically by 
the Nahm construction
\be
\Phit^{mn} (x^i) = \int^{v/2}_{-v/2} ds \; w_m^\dag(s,x^i) s w_n(s,x^i),
\ee
\be
A_i^{mn}(x^i) = -i \int^{v/2}_{-v/2} ds\;   w_m^\dag(s,x^i) \del_i w_n(s,x^i),
\ee
where $w_m(s,x^i)$ are normalizable solutions to a certain linear differential 
equation in $s$ with coefficients constructed from the Nahm data and $x^i$. 
The relation between a class of
non-abelian self-dual strings and BPS monopoles in four dimensions
revealed in this paper suggests that one may be able to generalize 
the Nahm construction for monopoles to 
the non-abelian self-dual strings. It is tempting to interpret the variable $s$ as
the dimension $x^5$.
On the other hand, the inclusion of a constant NSNS $B$-field
in the BPS monopole simple amounts to a constant shift in the Nahm equation. 
The availability of the generalized Nahm construction may allow one to include
a constant $C$-field in a simple manner and check against the quantum geometry 
\cite{qg,CG} 
obtained from other pictures of the system.
We leave this interesting issue for future discussions.

\section*{Acknowledgements}

It is a pleasure to thank Ed Corrigan, Pei-Ming Ho, 
Hotoshi Isono, Sheng-Lan Ko, Kimyeong Lee,  
Douglas Smith, 
Pichet Vanichchapongjaroen and Martin Wolf 
for discussions. This work is supported in part 
by the STFC Consolidated Grant ST/J000426/1 and 
by the grant 101-2112-M-007-021-MY3
of the National Science Council, Taiwan.


\end{document}